\documentclass{aastex}
\usepackage{emulateapj5}
\usepackage{onecolfloat5}

\newcommand{\gsim}{\;\rlap{\lower 2.5pt
 \hbox{$\sim$}}\raise 1.5pt\hbox{$>$}\;}
\newcommand{\lsim}{\;\rlap{\lower 2.5pt
   \hbox{$\sim$}}\raise 1.5pt\hbox{$<$}\;}
\newcommand{\be}{\begin{equation}}
\newcommand{\beq}{\begin{equation}}
\newcommand{\ba}{\begin{eqnarray}}
\newcommand{\ee}{\end{equation}}
\newcommand{\eeq}{\end{equation}}
\newcommand{\ea}{\end{eqnarray}}

\newcommand{\cmb}{\Theta}
\def\vecl{{\bf l}}
\newcommand{\tot}{{\rm tot}}

\newcommand{\bea}{\begin{eqnarray}}
\newcommand{\eea}{\end{eqnarray}}
\newcommand{\bean}{\begin{eqnarray*}}
\newcommand{\eean}{\end{eqnarray*}}
\newcommand{\bq}{{\bf q}}
\newcommand{\bx}{{\bf x}}
\newcommand{\bk}{{\bf k}}
\newcommand{\bn}{{\bf \hat{n}}}
\newcommand{\dxe}{{\delta_{X_e}}}

\begin{document}

\twocolumn[
\title{Small-scale CMB Temperature and Polarization Anisotropies Due to
Patchy Reionization}
\author{M\'ario G. Santos}
\affil{Department of Physics, University of California, Davis\\ Email: santos@physics.ucdavis.edu}
\author{Asantha Cooray}
\affil{Theoretical Astrophysics, MS 130-33, California Institute of Technology, Pasadena, CA 91125.\\E-mail: asante@caltech.edu}
\author{Zolt\'an Haiman}
\affil{Department of Astronomy, Columbia University, 550 West 120th
Street, New York, NY 10027}
\author{Lloyd Knox}
\affil{Department of Physics, University of California, Davis}
\author{Chung-Pei Ma}
\affil{Department of Astronomy, University of California, Berkeley}

\begin{abstract}
We study contributions from inhomogeneous (patchy) reionization
to arcminute scale ($1000 < \ell < 10,000$) cosmic microwave background
(CMB) anisotropies. We show that inhomogeneities in the ionization fraction,
rather than in the mean density, dominate both the temperature
and the polarization power spectra.
Depending on the ionization history and the clustering bias of the ionizing
sources, we find that rms temperature fluctuations range from 2 $\mu$K to 
8 $\mu$K and the corresponding values for polarization are over two orders of magnitude
smaller.  Reionization can significantly
bias cosmological parameter estimates and degrade gravitational lensing
potential reconstruction from temperature maps but not from polarization
maps.  We demonstrate that a simple modeling of the reionization temperature
power spectrum may be sufficient to 
remove the parameter bias.  The high-$\ell$ temperature power spectrum will
contain some limited information about the sources of reionization.\\
\end{abstract}]

\section{Introduction}

Free electrons in the reionized (or reionizing) intergalactic medium
(IGM) with peculiar velocities lead to new anisotropies as the CMB
photons scatter off of them.  Here we revisit predictions for the
temperature and polarization power spectra of these new anisotropies
at small angular scales in light of recent indications from
observations by the Wilkinson Microwave Anisotropy Probe (WMAP) for an
extended period of reionization \citep{kogut03}.

The argument for an extended period of reionization is as follows.
The WMAP has detected the correlation between temperature and polarization
on large angular scales \citep{kogut03} that has an amplitude proportional
to the 
total optical depth of CMB photons to Thomson scattering, $\tau$
\citep{sunyaev80,
zaldarriaga97a,kaplinghat02}.  Modeling
reionization with a single sharp transition at $z_{ri}$, a
multi--parameter fit to the WMAP data gives $z_{ri} = 17 \pm 5$
\citep{spergel03}.
On the other hand, the evolution of quasar spectra from $z=6.3$ and
$z=6.4$
to $z = 6$ shows a rapid decrease in the amount of neutral Hydrogen,
indicating
the end of reionization \citep{fan03}.  A simple
interpretation to explain these two very different datasets is that
reionization started early, $z_{ri} \sim 20$, but did not conclude
until much later ($z \sim 6$). 

As detected by  WMAP, free electrons in the IGM scatter
CMB photons and produce large angular scale CMB polarization. The
fluctuations in the electron density field lead to small angular scale
signatures both in the temperature and polarization. In the case of
temperature, the new small scale anisotropy contribution is  generally
known as the kinetic Sunyaev-Zel'dovich (kSZ) effect \citep{sunyaev80}
Inhomogeneities in the electron density may be due
to perturbations in the baryon density or ionization fraction.  If
the ionized fraction is taken to be homogeneous and baryon density perturbations are
calculated to first order then the effect is called the
Ostriker-Vishniac (OV) effect \citep{ostriker86,vishniac87}. Non-linear effects
from density perturbations make additional contributions beyond the OV effect on
small scales \citep{hu00b,gnedin01,ma02}. When
only modulations in the ionized fraction are considered then we refer to the
resulting temperature power spectrum as the patchy power spectrum.  We
focus on the patchiness because we find that, for the models we
consider, variations in the ionized fraction are much more important than variations
in the baryon density.  For a review of the imprint of reionization on
the CMB see \citet{haiman99}.

This extended period of reionization may be one in which the
ionization fraction is highly inhomogeneous, i.e.``patchy''.  As
discussed by \citet{aghanim96}, \citet{gruzinov98} and \citet{knox98}
(hereafter K98) such patchy reionization can lead to much larger anisotropies
than homogeneous reionization (see also \citealt{valageas01}, \citealt{aghanim02}).  
\citet{aghanim96} and
\citet{gruzinov98} modeled the patches as spatially uncorrelated.  K98
considered a model of correlated patches, and showed that these
correlations were very important because they can greatly increase the
signal in the $\ell = 1000$ to 3000 range.  In this paper we will take
into account the correlations between these reionized patches by
assuming they are associated with ionizing sources residing in dark
matter halos. The correlations will then be proportional to the linear
theory dark matter correlations, with the proportionality constant
given by an average of the halo bias weighted by halo mass and
an ionizing photon production efficiency factor,
since these determine the size of the individual ionized
patch around each dark halo.  In our models, the kSZ power spectrum
typically increases by an order of magnitude when the inhomogeneities in
the ionized fraction are taken into account.

Patchy reionization also boosts the creation of polarization anisotropies
on small angular scales.
K98 pointed out that this polarization signal,
since it is sourced by the temperature quadrupole as opposed to the
velocities which have grown by gravitational instability, would be
much smaller than the temperature signal.  This was subsequently
confirmed by explicit calculation \citep{weller99,hu00b}.
\citet{hu00b} pointed out that the B--mode power spectrum would be
equal to the E--mode power spectrum (for a flat-sky version,
see \citealt{baumann02}).  We also calculate the polarization
power spectrum from reionization and find that it is indeed too
small to significantly affect cosmological parameter determination or
lensing potential reconstruction.

Our paper is organized as follows.  In \S~\ref{sec:ksz} we describe the
kinetic SZ effect and how to calculate the temperature and
polarization angular power spectra in the presence of ionization
fraction and baryon density fluctuations.  In \S~\ref{sec:reion} we
briefly describe our models for reionization (described with more detail in
\citealt{haiman03}), including our calculation of the effective bias of
the patches.  In \S~\ref{sec:power} we discuss the resulting power spectra
and how they can be used as probes of reionization.
We also derive in \S~\ref{sec:measur} the expected errors from a 
ground--based experiment capable of detecting this signal. In \S~\ref{sec:bias} we
calculate the resulting bias in the cosmological parameter
determination from Planck data that can be caused by patchy
reionization, and show how simple phenomenological modeling can eliminate
these biases. Finally, in \S~\ref{sec:lens}, we discuss the impact on
lensing potential reconstruction. We conclude with a summary in \S~\ref{sec:concl}.
Throughout the paper, unless stated
otherwise, we will assume the cosmological parameters,
$\Omega_m=0.29$, $\Omega_\Lambda=0.71$, $\Omega_b=0.047$, $h=0.72$,
$\tau=0.17$ and $n=1$ with normalization $\sigma_8=0.9$ in agreement
with a recent analysis of CMB data \citep{spergel03}.

\section{Kinetic SZ effect}
\label{sec:ksz}

Compton-scattering via moving electrons leads to a Dop\-pler effect and
a corresponding temperature perturbation along the line-of-sight:
\be
\frac{\Delta T}{T}(\bn)=\int d\eta\,a(\eta)\,g(\bx,\eta)\,\bn \cdot {\bf v} \, ,
\ee
where $\bn$ is the unit vector along the direction of observation,
$\eta$ is the conformal time (with value $\eta_0$ today), $\bx=\bn(\eta_0-\eta)$ and ${\bf v}$ is the electron 
velocity. The visibility function, $g(\bx,\eta)$, corresponds to the probability that 
the photon scattered at time $\eta$ and traveled freely since then:
\be
g(\bx,\eta)=\sigma_T\,n_e(\bx,\eta)\,e^{-\tau(\eta)},
\ee
where $\tau$ is the optical depth to electron scattering between here
and distance $\eta_0-\eta$ and $\sigma_T$ is
the Thomson cross section.
The free electron density is $n_e=X_e(\bx,\eta) n_p(\bx,\eta)$ and both the 
ionization fraction, $X_e$ and proton density, $n_p$, are allowed to have 
a spatial dependence.

Under a smooth distribution of free electrons with $n_e=\bar{n}_e$, the 
Doppler contribution to temperature anisotropies is significantly reduced
as photons scatter against the crests and troughs of the velocity perturbations
\citep{kaiser84}. This cancellation is avoided when there are significant
fluctuations in the free electron density field, which can originate from
perturbations in the ionized fraction ($\delta_{X_e}$) or the 
baryon density ($\delta_b$).
The total contribution to the temperature anisotropy, up to second order,
can then be expressed as:
\be
\frac{\Delta T}{T}(\bn)= \sigma_T\,\bar{n}_p(\eta_0)
\int d\eta\,\left[a^{-2}(\eta) e^{-\tau} \bar{X}_e(\eta)\right]\,\,\bn \cdot {\bf q} \, ,
\ee
with 
\be
{\bf q}=(1+\delta_{X_e}+\delta_b)\, {\bf v}
\ee and
$\sigma_T\,\bar{n}_p(\eta_0) = 2.03\times 10^{-5}\,\Omega_b h^2\,{\rm Mpc}^{-1}$.

\subsection{Power spectrum}
We next calculate the kSZ power spectrum on small angular scales, where 
the signal is not overwhelmed by the primary anisotropies.
Then, for $\ell\gg 1$, we can use the Limber approximation \citep{limber54} and
write the spherical Fourier coefficient $C_\ell$ in terms of the 3D power spectrum of $\bq(\bx,\eta)$ 
\citep{kaiser92,dodelson95,hu96,jaffe98}:
\be
\label{Cl}
C_\ell=\sigma_T^2\,\bar{n}_p(\eta_0)^2 \int {d\eta\over x^2}\,
a^{-4}(\eta) e^{-2\tau} \bar{X}_e^2(\eta)\,P_{q_\perp}\left({\ell\over x},\eta\right),
\ee
where
\be
\langle \bq_\perp(\bk)\cdot\bq_\perp^*(\bk')\rangle =  
2(2\pi)^3 \delta(\bk-\bk')\,P_{q_\perp}(k)
\ee
and $\bq_\perp(\bk)$ is the component of the 3D Fourier transform of $\bq(\bx)$
that is perpendicular to $\bk$ ($\bq_\perp(\bk)\cdot \bk=0$).
The implicit assumption here is that the quantity 
$\left[a^{-2}(\eta) e^{-\tau} \bar{X}_e(\eta)\right]$ is slowly-varying across a wavelength of
the perturbation ($\Delta\eta\sim \eta_0/\ell$ at most).  
This is a good approximation for the angular scales of interest
($\ell\gsim 1000$).

The key quantity to compute is therefore $P_{q_\perp}$ for all possible
contributions. To first order $\bq={\bf v}$ and ${\bf v}(\bk)\propto \bk$ so that,
as already stated, there will be no contribution to the CMB power spectrum from
equation (\ref{Cl}).
At larger angular scales, on the other hand, the anisotropy from the component
of $\bq(\bk)$ parallel to $\bk$ becomes significant and is taken into account in
standard Boltzmann calculations.
Considering the density--modulated and ionized--fraction--modulated anisotropies,
the components of $\bq(\bk)$ perpendicular to $\bk$ are
\bea
q_{\perp,i}(\bk)&=&-i\int {d^3k'\over (2\pi)^3}V(\bk')\left({k_i'\over k'}-\mu'{k_i\over k}\right)
\times\\ \nonumber
& &\left[\delta_b(|\bk-\bk'|)+\delta_{X_e}(|\bk-\bk'|)\right],
\eea
where $\mu'\equiv\hat{\bk}\cdot\hat{\bk}'$ and we have assumed that the velocity
field is a potential flow, ${\bf v}(\bk)=-i V(\bk)\hat{\bk}$, as expected
in linear perturbation theory.
The corresponding power spectrum is then
\bea
\label{tpower}
\nonumber
P_{q_\perp}(k)&=&{1\over2}\sum_{a,b=\delta_b,\delta_{X_e}}\int {d^3k'\over (2\pi)^3}
\left[{\over}(1-\mu'^2)P_{ab}(|\bk-\bk'|)P_{vv}(k')\right.\\ 
& &-\left.{(1-\mu'^2)k'\over |\bk-\bk'|}
P_{av}(|\bk-\bk'|)P_{bv}(k')\right].
\eea
Note that, due to geometric reasons, there is no
connected fourth moment contribution to $P_{q_\perp}(k)$ \citep{cooray01,ma02}.

\subsection{Density-modulated anisotropies}

If reionization was homogeneous throughout the Universe, temperature
anisotropies can only arise from perturbations in
the baryon density (e.g. $a=b=\delta_b$ in equation \ref{tpower}). When
$\delta_b$ is in the linear regime, this is the Ostriker-Vishniac effect
\citep{ostriker86,vishniac87}, which has already been studied in some detail 
\citep{dodelson95,hu94,persi95,hu96,jaffe98,hu00b}. Here we will just try to give an order of magnitude estimate
for the effect. For wavelengths that are much smaller than the coherence
scale of the velocity fields, equation (\ref{tpower}) simplifies considerably.
First we note that, using the continuity equation:
\bea
\nonumber
P_{vv}(k',\eta)\propto P_{\delta_c\delta_c}(k')/k'^2
\,\,\,\,\, 
P_{\delta_cv}(k',\eta)\propto P_{\delta_c\delta_c}(k')/k',
\eea
where we have assumed that the baryonic gas traces the cold dark
matter. For large $k'$, typically larger than $k_{m}\sim 0.1\ h/$Mpc
\citep{hu00b} we then have
$P_{vv}\sim 0$ and $P_{\delta_c v}\sim 0$ (density fluctuations
uncorrelated with the bulk velocity field). Therefore, for large $k$
the integral in equation (\ref{tpower}) can be evaluated under the approximation
$|\bk-\bk'|\approx k$ so that,
\be
\label{spower}
P_{q_\perp}(k,\eta)\sim {1\over 3}P_{\delta_c\delta_c}(k,\eta)v_{rms}^2(\eta),
\ee
with 
\be
v_{rms}^2=\int{k^2 dk\over 2\pi^2} P_{vv}(k,\eta).
\ee
The factor of 1/3 arises from angular averaging, since only the line 
of sight component of the velocity field contributes to the signal
(the two transverse parts drop out).
For these small wavelengths, typically $P_{\delta_c\delta_c}(k)\propto k^{-3}$ so that,
introducing the above expression in equation (\ref{Cl}), gives 
$C_\ell\propto \ell^{-3}$ (for $\ell\gg \eta_0\,k_m$). Moreover, assuming our LCDM
fiducial model and instantaneous reionization of the Universe, we obtain,
as an order of magnitude estimate,
$\ell^2C_\ell/(2\pi) \sim 3\times10^{-13}(5000/\ell)$.

\begin{figure}[!bt]
\vspace{-0.5cm}
\plotone{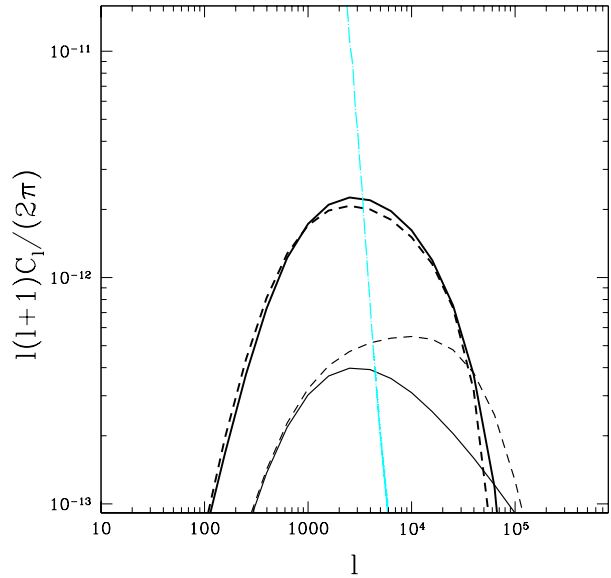}
\vspace{-0.6cm}
\caption{The several contributions to the kSZ temperature power
spectrum assuming the LCDM model
described in the introduction. The thin solid line is the O-V power spectrum
from homogeneous reionization and dashed line is the non-linear correction.
The thick solid line is the typical contribution from
inhomogeneous reionization and uses the model shown in Figure~\protect\ref{fig:xe} and
\protect\ref{fig:bias} (solid line). Dashed line shows the approximate model from eq.
(\protect\ref{simpower}). The patchy power spectrum basically follows the shape of the
matter power spectrum. Also shown is the primary lensed CMB power spectrum (dot-dashed).}
\label{fig:ksz} 
\end{figure}

In the low redshift universe most of the 
baryonic matter resides in collapsed, high density
objects. We therefore need to modify the above expressions to take into account the
non-linearities of the density perturbations \citep{hu00b,gnedin01,ma02}. These non-linearities only affect the
density field below the coherence scale of the bulk velocity. Therefore, 
equation (\ref{spower}) will be a good approximation for scales that become non-linear.
One then only needs to obtain the corresponding non-linear matter power spectrum using an appropriate
prescription such as the halo model \citep{ma00}. Figure \ref{fig:ksz} compares 
the OV power spectrum with the non-linear correction calculated using the halo model
(thin lines).
We see that for sufficiently large $\ell$ there is an increase in power due to a
similar increase in the matter power spectrum.
We point out that this non--linear OV effect includes contributions to the kSZ signal 
from low redshift galaxy clusters. As will be seen below, the amplitude of the kSZ signal from the 
high--redshift ionized patches overwhelms this low--redshift contribution in most reionization
models. The results in \citet{ma02} agree reasonably well with hydrodynamical simulations
\citep{springel01,silva01,white02} for the angular scales of interest.

\subsection{Patchy reionization}

Exactly as with perturbations in the matter density, 
inhomogeneities in the ionized fraction create temperature
anisotropies through modulation of the Doppler effect (K98, \citealt{gruzinov98}). 
Considering the case where $a=b=\delta_{X_e}$ in equation (\ref{tpower}),
we now need to obtain an expression for $P_{\dxe\dxe}(k)$.
We assume the ionizing radiation is coming from stars formed from gas clouds
that cooled in dark matter halos.
The dark halo correlation function can be expressed in terms of the 
linear theory dark matter one \citep{mo96,jing98}:
\be
\xi_{halo}(r;M,M',\eta)=b(M,\eta)b(M',\eta)\xi_{\delta_c\delta_c}(r,\eta),
\ee
where $\xi_{halo}(r;M,M')$ is the correlation function between
halos of mass $M$ and $M'$ and $b()$ is the clustering bias parameter. 
To obtain the clustering properties of the ionized fraction we need to
take into account the dependence of the size and spatial distribution
of the ionizing patches on the masses
and ionization efficiencies of dark halos. Therefore, one should be able to
write the correlation as:
\be
\xi_{\dxe\dxe}(r,\eta)=b^2_{eff}(\eta)\xi_{\delta_c\delta_c}(r,\eta)
\ee
where $b_{eff}$ is the mean bias weighted by the different halo properties.
Since these ionizing patches are rare objects we expect them to be more clustered than the
underlying dark matter (e.g. $b_{eff}>1$). We will compute this bias in more detail in
Section \ref{sec:reion}.
Finally, one has to take into account that the perturbations will be smoothed
out on scales smaller than the size of the patch. Choosing a Gaussian filter,
the  power spectrum for the ionized fraction will then be
\be
\label{Pxexe}
P_{\dxe\dxe}(k,\eta)=b^2_{eff}(\eta)P_{\delta_c\delta_c}(k,\eta)e^{-k^2R^2},
\ee
where $R$ is the mean radius of the ionized patches (HII regions).

We crudely model the time--dependence of $R$ as
\be
\label{eqn:Rpatch}
R\sim\left({1\over 1-\bar{X}_e}\right)^{1/3}R_p,
\ee
where $R_p$ is the (comoving) size of the fundamental patch which
we take to be $R_p\sim 100$\, Kpc.   The cutoff scale $R$ is equal to
$R_p$ when reionization starts and increases with time as HII regions overlap and form
larger HII regions.  Equation~\ref{eqn:Rpatch} follows from assuming
a Poisson distribution of patches and that their volumes add when they
overlap.  But the essential consequence of equation~(\ref{eqn:Rpatch}) is
that on the scales of interest ($\ell \la 10,000$ or $R \ga 1$Mpc) the
transition is very sudden; i.e., $R$ increases from 1 Mpc to $\infty$
very rapidly.  We expect this behavior to be generic and thus
our results should be insensitive to the specifics of the overlap modeling. 

To calculate the patchy power spectrum we neglect the cross--correlation
between velocity and $\delta_{X_e}$ since it is small, though we 
discuss this contribution later.  Then, 
we simply plug expression (\ref{Pxexe}) into equation~(\ref{tpower}). 
The result for our fiducial reionization model (discussed in the next section)
is shown in Figure~\ref{fig:ksz}.
To have a better understanding of the expected shape and amplitude of
this power spectrum we will now derive an approximate expression. 
Using an expression 
similar to equation (\ref{spower}) we find the patchy power spectrum to be
\bea
\nonumber
C_\ell={\sigma_T^2\,\bar{n}_p(\eta_0)^2\over 3}& &\int_{\eta_i}^{\eta_1} {d\eta\over x^2}\,
a^{-4}(\eta) e^{-2\tau} \bar{X}_e^2(\eta)\times\\
& &P_{\dxe\dxe}\left({\ell\over x},\eta\right)v_{rms}^2(\eta),
\label{eqn:TkSZ}
\eea
where $\eta_i$ signals the start of reionization (when $X_e$ starts increasing from
its value at recombination) and $\eta_1$ corresponds to the time when
$\bar{X}_e=1$. We can write the time dependence of the dark matter perturbation as
$\delta_c(k,\eta)\ =\ \left({G\over G_0}\right)\delta_c(k,\eta_0)$,
where $G$ is the growth factor \citep{carroll92} and $G_0$ is the value today.
Moreover, assuming $x_i-x_1\ll x_1$, we can pull the $\ell$
dependence out of the time integration, so that:
\bea
\label{ppower}
\nonumber
C_\ell&\approx&{\sigma_T^2\,\bar{n}_p^2\over 3 \bar{x}^2}P_{\delta_c\delta_c}\left({\ell\over \bar{x}}\right)
e^{-{\ell^2 \bar{R}^2\over \bar{x}^2}}\left({v_{rms}(\eta_0)\over G_0 \dot{G}_0}\right)^2\times\\
& &e^{-2\bar{\tau}}\int_{\eta_i}^{\eta_1} d\eta\,a^{-2} \bar{X}_e^2(\eta)\,b_{eff}^2(\eta)
 (G\dot{G})^2,
\eea
where $\dot{G}\equiv {dG\over d\eta}$.
The quantities, $x$, $R$ and $\tau$ are assumed to be evaluated at the average time, 
$(\eta_1+\eta_i)/2$. 
At the redshifts of interest for the above time integration, the cosmological constant
is negligible, so that $G\sim a$ and $H\sim H_0\sqrt{\Omega_m/a^3}$.
The power spectrum can then be expressed as
\bea
\label{simpower}
\nonumber
\ell^2C_\ell/(2\pi)&\sim& 6.18\times 10^{-21}\ell^3\,T^2(\ell/\bar{x})
e^{-{\ell^2 \bar{R}^2\over \bar{x}^2}}\times\\
& &\int_{z_1}^{z_i}{dz\over \sqrt{1+z}} \bar{X}_e^2(z)\,b_{eff}^2(z),
\eea
where we are using our fiducial cosmological model, so that
$v_{rms}(\eta_0)/c=1.8\times 10^{-3}$, $\bar{x}\sim8.0\times 10^3$ Mpc/h and
$\bar{R}\sim0.20$ Mpc/h. The matter power spectrum is normalized to give
$\sigma_8= 0.9$ and the transfer function, $T(k)$ 
\citep{bardeen86,eisenstein98,seljak96}, is normalized to 1 on large
scales. All the dependence on the reionization
model is in the time integration, through $\bar{X}_e$ and $b_{eff}$. Therefore we expect the
power spectrum from different reionization histories to have approximately the same shape but
different amplitudes. Figure \ref{fig:ksz} shows a possible power spectrum from the above
patchy reionization model. Also shown is the approximate model from equation (\ref{simpower}).
This signal typically overwhelms the OV effect, even when including the
non-linear correction, which can be traced back to the large effective bias,
$b_{eff}\gsim 5$, in equation (\ref{ppower}).

We have ignored the term, $\langle v_1 \dxe_2 \rangle\langle v_2 \dxe_1 \rangle$ 
as sub-dominant to $\langle \dxe_1 \dxe_2 \rangle
\langle v_1  v_2 \rangle$.  The former term may become
important for length scales above the velocity coherence length, which
projects to $l \sim 500$ from high redshifts.  
Due to the tendency of overdense regions to fall toward each other, 
this term is negative and the net effect will be 
a suppression of power.  This power suppression will be small in the
scales of interest.  We have also neglected the cross correlation 
$\langle \delta_c\dxe\rangle$ because it will be suppressed by approximately
one factor of the bias from $\langle \dxe\dxe \rangle$.

\subsection{Polarization Signatures}

\begin{figure}[!bt]
\vspace{-0.2cm}
\plotone{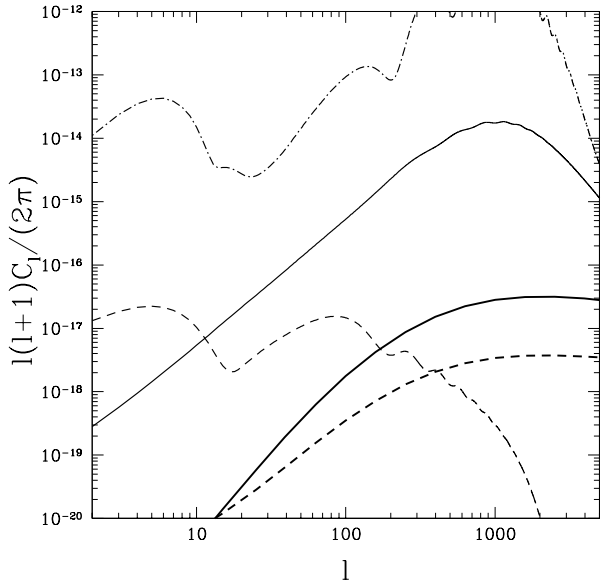}
\vspace{-0.6cm}
\caption{$B$ mode polarization for our fiducial model, with contributions from lensing
(thin solid line) and tensor modes (thin-dashed) with a tensor to scalar ratio of
$10^{-3}$. Also shown is the contribution from density (thick dashed) and ionization
(thick solid) modulated scattering. The density modulated contribution assumes gas
traces dark matter and uses the halo model for non-linear corrections. Also shown
for comparison is the scalar E mode contribution (dot-dashed line).} 
\label{fig:pol} 
\end{figure}

Thomson scattering of radiation with a quadrupole ani\-so\-tropy generates
linear polarization in the CMB. Again, the dominant source at small
angles is the modulation of the quadru\-pole through perturbations in the
matter density or ionized fraction. The power spectra of the $E$
and $B$ parity states are then \citep{hu00b,baumann02}
\bea
\nonumber
C_\ell^{EE}&=&C_\ell^{BB}={3\sigma_T^2\,\bar{n}_p(\eta_0)^2\over 100}\int {d\eta\over x^2}\,
a^{-4}(\eta) e^{-2\tau} \bar{X}_e^2(\eta)\times\\
& &\left(P_{\delta_c \delta_c}+P_{\delta_{X_e} \delta_{X_e}}\right)\, Q_{rms}^2(\eta),
\label{eqn:Pksz}
\eea
where $Q_{rms}$ is the quadrupole anisotropy. 
The power spectrum from the primordial quadrupole source will
typically dominate over the kinematic and intrinsic contributions \citep{hu00b}. Moreover,
this quadrupole is a slowly varying function of time and can be taken out of the
integral above. Figure \ref{fig:pol} shows the patchy and density contributions to
polarization (using $Q_{rms}\sim 25\mu$K), compared to the $B$ mode polarization
from tensor and lensing effects. Again the patchy signal dominates over the density
modulated polarization. 
However, it does not seem to be an important contaminant on the
relatively large scales  ($\ell\lsim 200$) where tensor $B$ modes 
are significant, and it is well below the
scalar $E$ mode on all scales.

The patchy polarization power spectrum is $\sim 10^5$ times smaller than
the patchy temperature power spectrum because $Q_{\rm rms} \sim 10^{-2} v_{\rm rms}$
and due to the geometric factor of 100 in equation~\ref{eqn:Pksz} in place of the
factor of 3 in equation~\ref{eqn:TkSZ}.

\section{Reionization Models}
\label{sec:reion}

We now need to obtain the evolution of the electron fraction, $\bar{X}_e$ and
effective bias, $b_{eff}$, in
physically motivated models of the reionization history. For that purpose, we
adapt semi--analytical models from \citet{haiman03}, modified
to include the spatial biasing of the ionizing sources and hence of
the ionized regions.  These models fit all the relevant observations
(including the electron scattering optical depth $\tau=0.17$ measured
by WMAP and inferences from redshift $z\sim 6$ quasar spectra).  For
details, the reader is referred to that paper; here we
provide only a brief summary, and a description of how the bias is
calculated.

We follow the volume filling fractions $F_{\rm HII}$ of HII regions
(which, by definition, equals the mean electron fraction $\bar{X}_e$)
assuming that discrete ionized Str\"omgren spheres are being driven
into the IGM by ionizing sources located in dark matter halos.  In
this picture, each fluid element is engulfed by an ionization front at
a different time.  
Reionization has a complex history that
reflects contributions from three distinct types of ionizing sources,
and two different feedback effects (all of which have physical
motivations as described in detail in \citealt{haiman03}).  In
short, ionizing sources (assumed to be massive metal--free stars)
first appear inside gas that cools via ${\rm H_2}$ lines, and collects
in the earliest non--linear halos with virial temperature of
$100\,{\rm K} \lsim T \lsim 10^4$K \citep{haiman96}.
These sources ionize $\sim50\%$ of the volume in hydrogen. However, at
this stage (redshift $z\sim 17$) the entire population of these
sources effectively shuts off due to global ${\rm
H_2}$--photodissociation by the cosmic soft UV background they had
built up \citep{haiman97a,haiman97b,haiman00}.  Soon
more massive halos, with virial temperatures of $10^4\,{\rm K} \lsim T
\lsim 2\times 10^5$K form, which do not rely on ${\rm H_2}$ to cool
their gas (they cool via neutral H excitations), and new ionizing
sources turn on in these halos.  These are assumed to be ``normal''
stars, since the gas had already been enriched by heavy elements from
the first generation.  This population continues ionizing hydrogen,
but is also self--limiting: gas infall to these relative shallow
potential wells is prohibited inside regions that had already been
ionized and photo--heated to $10^4$K \citep{thoul96}.  As a
result, hydrogen reionization starts slowing down around $z\sim
10$. However, at this stage, still larger halos with virial
temperatures of $T \gsim 2\times 10^5$K start forming. These
relatively massive halos are impervious to photoionization feedback,
and complete the reionization of hydrogen at around $z\approx 7$.

In order to predict the kSZ signal, we need to compute in our models the
mean bias of the ionized regions as a function of redshift.  At any
given redshift, there is an ensemble of HII regions, the sum of
which make up the total ionized volume fraction.  Each HII region
has at least one ionizing source, each of which has a fixed halo mass
and formation redshift.

\begin{figure}[!bt]
\vspace{-0.2cm}
\plotone{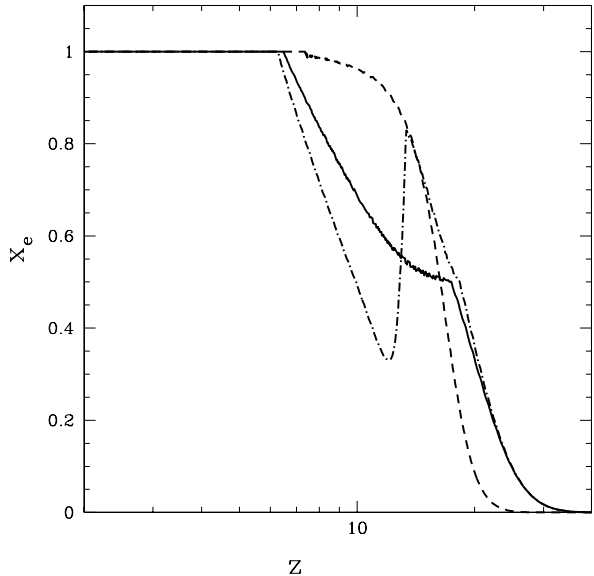}
\vspace{-0.6cm}
\caption{$\bar{X}_e$ for three different fiducial models with $\tau=0.17$.
Solid line
is for the fiducial model used throughout the paper. Dashed line assumes
that $H_2$
feedback effectively shuts off star formation in all Type II halos.
Dot-dashed line
assumes that a transition from metal free to normal star formation takes
place
abruptly at a certain redshift. 
\label{fig:xe} }
\end{figure}

We assume that the volume of each ionized region scales linearly with halo
mass within each of three distinct mass bins; we only allow a dependence
of ionizing photon production efficiencies in discrete jumps
between these bins.  This assumption could be relaxed to include
more generic dependencies on halo mass. This, in general, would
change the effective bias we obtain, but would involve introducing
more free parameters; we avoid this complication in the present paper.
Under these assumptions, the mean bias of ionized
regions can be computed as follows.  We first note that at a given
redshift, the total volume--filling fraction 
$F_{\rm HII}$ is given by

\begin{eqnarray}
\nonumber
F_{\rm HII}&&(z)=\rho_{\rm b}(z)
\int_{\infty}^{z}dz^{\prime} 
\left\{
  \epsilon_{\rm Ib} \frac{dF_{\rm coll,Ib}}{dz} (z^{\prime}) 
+ \left[1-F_{\rm HII}(z^\prime)\right]\right.\times\\
&&\left.\left[  \epsilon_{\rm Ia} \frac{dF_{\rm coll,Ia}}{dz} (z^{\prime}) 
        + \epsilon_{\rm II}  \frac{dF_{\rm coll,II}}{dz} (z^{\prime})
 \right]
\right\}
{\tilde V}_{\rm HII}(z^{\prime},z),
\label{eq:filling}
\end{eqnarray}
where $\rho_{\rm b}=\Omega_{\rm b}\rho_{\rm crit}$ is the average
baryonic density and $\rho_{\rm crit}$ is the critical density of the
universe.  In the second term on the right hand side of
equation~(\ref{eq:filling}), we have explicitly included a factor
$(1-F_{\rm HII})$, which takes into account the photoionization
feedback.  Here $F_{\rm
coll,II}$, $F_{\rm coll,Ia}$ and $F_{\rm coll,Ib}$, 
are the fractions of baryons collapsed into halos of
increasing mass ranges.  For example (using
$\bar \rho_{\rm tot}$ as the present--day total mass density),
\be
F_{\rm coll,II} \equiv {1 \over \bar \rho_{\rm tot}} \int_{M_{\rm
min}}^{M_{\rm max}} dM \frac{dn}{dM}(z) M
\label{eq:fcol}
\ee
where $M_{\rm min}$ and $M_{\rm max}$ correspond to halo virial
temperatures of $T_{\rm vir}=100$K and $T_{\rm vir}=10^4$K at redshift
$z$, respectively.
The $\epsilon_{\rm II}$, $\epsilon_{\rm Ia}$ and
$\epsilon_{\rm Ib}$ are the efficiencies in these halos of producing
and leaking ionizing photons into the IGM. The efficiencies are
defined as the product $\epsilon_* \equiv N_\gamma f_* f_{\rm esc}$,
where $f_* \equiv M_*/(\Omega_{\rm b}M_{\rm halo}/\Omega_m)$ is the
fraction of baryons in the halo that turns into stars; $N_\gamma$ is
the mean number of ionizing photons produced by an atom cycled through
stars, averaged over the initial mass function (IMF) of the stars; and
$f_{\rm esc}$ is the fraction of these ionizing photons that escapes
into the IGM.  Finally, ${\tilde V}_{\rm HII}(z_{\rm on},z)$ is the
volume per unit mass and unit efficiency ionized by redshift $z$ by a
single source that turned on the earlier redshift $z_{\rm on}>z$.
The evolution of $F_{\rm HII}$ for a few models described in
\cite{haiman03} is shown in Figure~\ref{fig:xe}. They all produce a total electron 
scattering optical depth of
$\tau=0.17$ by construction, in agreement with the {\it WMAP} result.
For the one we take as our fiducial model in this paper (solid line),
the efficiencies are
$\epsilon_{\rm II}=200$ and $\epsilon_{\rm Ia}=\epsilon_{\rm Ib}=80$.

To calculate the effective clustering bias for any field, the
first step is to write the field as an integral over halo mass and
formation redshifts, which we have effectively done with 
equation~(\ref{eq:filling}) and (\ref{eq:fcol}).  
We then use this sum to define $b_{eff}$ as the appropriate 
weighted average of the bias factor for a halo of mass $M$ and
formation redshift $z$, $b(z,M)$.  
The effective bias for the ionization fraction for our 
reionization models can thus be efficiently written as:
\be
b_{eff}(z) = F^b_{\rm HII}/F_{\rm HII}
\label{eq:globalmeanbias}
\ee
where
\begin{eqnarray}
\nonumber
F^b_{\rm HII}&&(z)\equiv\rho_{\rm b}(z)
\int_{\infty}^{z}dz^{\prime} 
\biggl\{
  \epsilon_{\rm Ib}b_{\rm Ib}(z^\prime) \frac{dF_{\rm coll,Ib}}{dz}
(z^{\prime})+\\
\nonumber 
&& \left[  \epsilon_{\rm Ia}b_{\rm Ia}(z^\prime) \frac{dF_{\rm
coll,Ia}}{dz} (z^{\prime}) 
        + \epsilon_{\rm II}b_{\rm II}(z^\prime)  \frac{dF_{\rm
coll,II}}{dz} (z^{\prime})
 \right]\times\\
&& \left[1-F_{\rm HII}(z^\prime)\right]\biggr\} {\tilde V}_{\rm
HII}(z^{\prime},z)
\label{eq:fillingb}
\eea
and 
\be
b_{\rm II}(z)\equiv
\left[F_{\rm coll,II}\,\bar \rho_{\rm tot}\right]^{-1}
\int_{M_{\rm min}}^{M_{\rm max}} dM \frac{dn}{dM}(z) b(z,M)M,
\label{eq:meanhalobias}
\ee
with analogous expressions defining 
$b_{\rm Ia}(z)$ and $b_{\rm Ib}(z)$.

In this last equation, $b(z,M)$ is the usual halo 
bias (Mo \& White 1996). The averaging over formation redshift
only introduces a small correction to the bias calculation, 
since at a given redshift most of the contribution comes from 
young ionizing sources.

Figure~\ref{fig:bias} shows the effective bias for the same three
models as in Figure~\ref{fig:xe}. The increase of the bias as $z$
decreases toward $\sim 7$ is due to the formation of rare, massive
halos, which play an important role in the last stages of
reionization.

Note that the bias factor as calculated in Eq.~(\ref{eq:globalmeanbias}), 
Eq.~(\ref{eq:fillingb}) and
(\ref{eq:meanhalobias}) agrees with the halo
model description of source bias \citep{cooray02} when one
makes the  connection between the presence of ionized electrons and
the halo occupation distribution. Here, the bias factors are calculated
with the assumption that ionized electrons occupy halos with mass in the
range $M_{\rm min}$ and $M_{\rm max}$, such that the mean halo occupation
number is unity in this mass range and zero otherwise.

\begin{figure}[!bt]
\vspace{-0.2cm}
\plotone{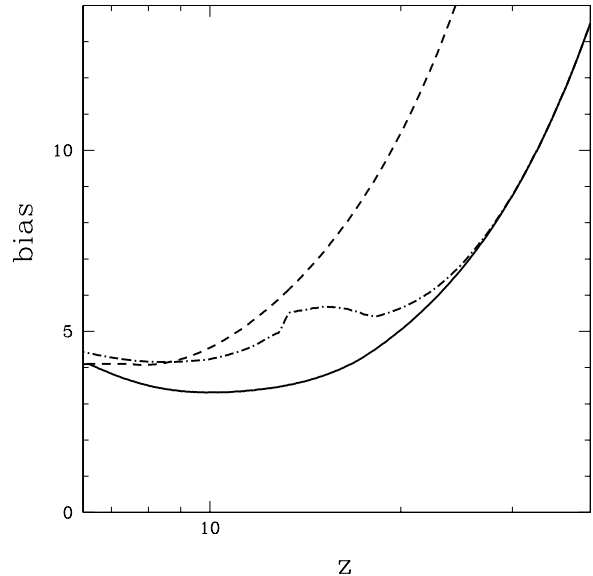}
\vspace{-0.6cm}
\caption{Effective bias (equation \ref{eq:globalmeanbias})
for the same fiducial models as in Figure~\protect\ref{fig:xe}.} 
\label{fig:bias} 
\end{figure}

We now remark on the robustness of the model of patchy
reionization we have presented.  This simple picture has the HII
regions localized around the halos hosting the ionizing sources.
Further, our calculation of $V(z,z')$ assumes the recombination rate
is given by the square of the background density times some constant
clumping factor.  In reality, there must be complicated shapes for the
volume
$V(z,z')$, which depend on the three-dimensional arrangement of the 
density field, and also
on the locations and strengths of the ionizing sources.
The degree of localization of the HII regions around the sources
of ionizing photons will also depend on how hard the radiation
is, with harder spectra reionizing more uniformly
\citep{oh01,venkatesan01}.
And, if reionization is from decaying dark matter then $X_e$
would be yet more uniform \citep{hansen03}.

Another simple, but complementary, 
picture \citep{miralda-escude00} (hereafter MHR) is
that the ionizing radiation from a source leaves the surrounding dense
regions neutral and the i-fronts propagate along directions of the the
lowest column density into the voids.  In this case it actually does
not matter that the sources are in high-$\sigma$ peaks.  MHR postulate
that all gas below some threshold density, $\rho_t$, is ionized, and
all gas above this $\rho_t$ remains neutral.  The patchy power
spectrum in this case could be computed using the two-point function
of regions with $\rho<\rho_t$.

The MHR picture is the extreme opposite of ours (see also
\citealt{peebles98}). Which picture is a 
better description of reality depends entirely
on how rapidly percolation occurs.  We assume it only happens as
$\bar X_e$ gets very close to unity.  For our calculations to be
correct at $l \la 10,000$ requires that prior to $\bar X_e \sim 1$,
the ionizing radiation in the HII regions mostly comes from 
sources closer than 3 Mpc.

\section{Temperature Power Spectra}
\label{sec:power}

In Figure~\ref{fig:errors}, we plot the patchy power spectrum for the
three reionization models (including our fiducial model), all with the
same cosmological parameters and $\tau=0.17$ but differing $X_e(z)$
and $b_{\rm eff}(z)$.  As expected, the shapes are very
similar. Allowing $\tau$ to vary also does not significantly affect
the shape, as demonstrated by the fourth model plotted with
$\tau=0.11$ but higher bias that is close to a
$\tau=0.17$ model.  The higher bias is achieved by 
moving the ionizing sources to more massive halos (Type Ia and Ib).
The optical depth is prevented from dropping 
even further by increasing the efficiencies in Type Ia halos.

\begin{figure}[!bt]
\vspace{-0.2cm}
\plotone{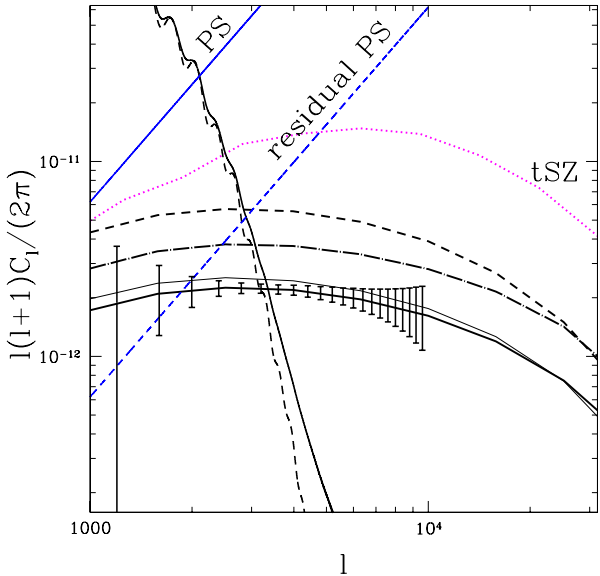}
\vspace{-0.6cm}
\caption{Patchy power spectra for the reionization models in
Figure~\protect\ref{fig:xe} (same line styles), together with other
astrophysical contributions and expected measurement errors (see
text).  The solid (dashed) straight line is the the point source
contribution at 217 GHz before (after) multi-frequency cleaning.  The
primary unlensed (dashed) and lensed (solid) CMB power spectra are
also shown as is the thermal SZ power spectrum from \citet{white02}
(dotted) with its expected amplitude at lower frequencies.  The thin
line close to the solid one shows the patchy power
spectrum for a model with $\tau=0.11$ but large bias.}
\label{fig:errors} 
\end{figure}

The degeneracy between $\tau$ and bias means that $\tau$ cannot be
inferred from the kSZ power spectrum alone, as suggested by
\citet{zhang03}.  However, if $\tau$ becomes tightly constrained by
polarization measurements at low $\ell$, which is possible with future
missions \citep{holder03}, the mean effective bias for the sources
of reionization can be determined.
Having this information will constrain the typical halo masses
of the sources that dominate reionization.  Whether the sources
are located in the 2-sigma or in the 3-sigma peaks is important
for the physics of reionization (see \citealt{rees99}). This 
will also constrain the mean efficiency $\epsilon$.

To demonstrate the range of possible power spectrum amplitudes we
calculate the patchy power spectra for a variety of reionization models with values of
$\tau$ spanning the 2$\sigma$ range allowed by WMAP. In
Figure~\ref{fig:models}
we plot the patchy power spectrum with variations in $\tau$ about two
of the models shown in Figure~\ref{fig:errors} (solid and dashed lines).
Variations in $\tau$ for the solid (dashed) line are done by essentially
changing the efficiencies in the Type II (Ia) halos.
Again the shape is very similar (at least up to $\ell\sim 10^4$) with the
amplitudes ranging from $1\times 10^{-12}$ to $8\times 10^{-12}$ at
$\ell\sim 2000$.

At $\ell > 10^4$ the power spectrum is sensitive to the structure of the
reionized patches on comoving length scales smaller than 1 Mpc.  At these
smaller scales the details of the patch geometries and sizes become
important.  With sufficiently reliable modeling of the power spectrum
on these scales (which we have not attempted) observations at $l > 10^4$
could be used to constrain these patch properties, although contamination
by point sources will make such observations difficult.

\begin{figure}[!bt]
\vspace{-0.2cm}
\plotone{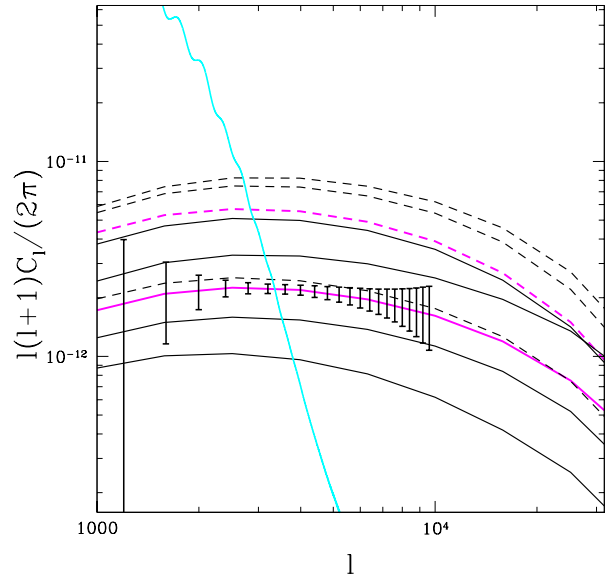}
\vspace{-0.6cm}
\caption{Patchy power spectrum for variations in $\tau$ about two of the models shown
in Figure~\ref{fig:errors} (thick solid and dashed lines). Solid lines correspond to
the following values of $\tau$ (from bottom to top): 0.05, 0.11, 0.17, 0.24, 0.31. Dashed
lines correspond to $\tau=$0.11, 0.17, 0.22, 0.28. Also shown are the expected measurement
errors and primary lensed CMB power spectrum.}
\label{fig:models} 
\end{figure}

\subsection{Measuring the kSZ Power Spectrum}
\label{sec:measur}

Ground-based observations, with 1$'$ to 2$'$ angular resolution, may be able to
measure the kSZ power spectrum.  In order to do so,
several other astrophysical signals must be cleaned from the maps:
namely, thermal SZ and emission from dusty galaxies. 

We take as an illustrative example a multi-frequency experiment that
includes a channel at 217 GHz.  We assume that the thermal SZ in the
217 GHz channel is insignificant, or at least that the residuals after
multi-frequency cleaning are insignificant.  We further assume the 217
GHz channel has an angular resolution of $\theta_b = 0'.9$ 
and an rms error of $\Delta_T = 12
\mu$K on every beam-size pixel in a fraction, $f_{\rm sky} = 0.01$,
of the sky, or 400 sq. degrees.

For our point source model, we adopt model E of \citet{guiderdoni98}.  For
this model, with 0$'$.9 resolution, 10 mJy point sources stand out as
5$\sigma$ fluctuations where $\sigma^2$ is the variance due to
sources fainter than 10 mJy.  Removing these bright sources with a
5$\sigma$ cut leaves a power spectrum due to the 
faint point sources of 
$l^2C^{\rm ps}_l/(2\pi)=2.5 \times 10^{-12}(l/2000)^2$ and
is plotted as the solid straight line in Figure~\ref{fig:errors}. 
Exploiting the frequency dependence of these remaining sources (with
respect to CMB temperature fluctuations) might reduce the power in
this contamination by a factor of 10 (dashed blue line).

The error on the total fluctuation power is
\bea
\Delta C_l^{\rm tot}& =& \sqrt{2 \over (2l+1)f_{\rm sky}}(C_l^{\rm
tot}+N_l) \, {\rm where} \nonumber\\
C_l^{\rm tot} &=& C_l^{\rm P}+C_l^{\rm PS}+C_l^{\rm kSZ}{\rm\ \ \ and}\nonumber\\
N_l &=& (\Delta_T \theta_b)^2\exp(l^2\theta_b^2/(8\ln{2})).
\label{eq:cltot}
\eea
Here $C_l^{\rm P}$ corresponds to the primary anisotropy contribution
and $C_l^{\rm kSZ}$ is the power spectrum from the kSZ effect.
A multi-parameter fit to $C_l^{\rm kSZ}$, the amplitude of a shot-noise
spectrum, and the cosmological parameters governing $C_l^{\rm P}$ will
result in
\be
\Delta C_l^{\rm kSZ} \approx \Delta C_l^{\rm tot}
\ee
since $C_l^{\rm P}$ and $C_l^{\rm PS}$ will be very well determined.
We plot the $\Delta C_l^{\rm tot}$ errors in Figure~\ref{fig:errors}
about our fiducial kSZ model.

\section{Impact on Parameter Determination}
\label{sec:bias}

The large anisotropies at $\ell \gsim 1000$ predicted from a patchy reionization phase may
significantly affect the high precision parameter estimation predicted for future
CMB experiments. 
Let us suppose we obtain a set of best fit parameters ${a_i}$ by minimizing the
likelihood, $\mathcal{L}^o$, without considering the patchy reionization signal
(e. g. ${\partial\ln{\mathcal{L}^o}\over\partial a_i}=0$). The above set should be
shifted by an amount $\delta a_i$ so that they minimize the true likelihood,
$\mathcal{L}^n$, which includes the patchy contribution:
\be
\delta a_i=\sum_{j}\mathcal{F}_{ij}^{-1}{\partial\ln{\mathcal{L}^n}\over\partial a_{j}},
\ee
where $\mathcal{F}_{ii'}$ is the fisher matrix,
\be
\mathcal{F}_{ij}=-\langle {\partial^2\ln{\mathcal{L}^n}\over\partial a_i\partial a_{j}}\rangle.
\ee
The bias in the estimation of each parameter will then be, as in K98,
\be
\delta a_i\approx -\sum_{j \ell} \mathcal{F}_{ij}^{-1} {C_\ell^R\over\sigma_\ell^2}
{\partial C_\ell^P\over\partial a_j},
\ee
where $C_\ell^R$ can represent the power spectrum from inhomogeneous reionization
and $\sigma_\ell$ is the error expected on the $C_\ell$'s, e.g. 
the sum of the errors due to sample variance
and noise, $N_\ell$ \citep{knox95,bond97,zaldarriaga97a}, $\sqrt{2\over 2\ell+1}(C_\ell^P+C_\ell^R+N_\ell)$.
We estimated the bias for the same set of cosmological parameters as in \citet{kaplinghat03a} taking 
into account the weak lensing contribution on small scales. 
Using our fiducial model for $C_\ell^R$, Table \ref{tab:planck}
shows the expected bias for Planck \citep{tauber01}. 

\begin{table*}[!bt]\small
\begin{center}
\caption{Bias and standard deviation in the parameter estimation for Planck ($\ell_{\rm max}=3000$).
\label{tab:planck}}
\begin{tabular}{l|rrrrrrrrrr}
    & $\Omega_m h^2$ & $\Omega_b h^2$ & $n_s$ & $n_s'$ & $m_\nu$ & $Y_{H_e}$ & $z_r$ &
$\theta_s$ & $\ln{P_\Phi}$ & $w_x$ \\
\tableline
bias    & 0.00012 & 0.00025 & -0.030 & -0.022 & -1.20 & -0.10 & 1.9 & -0.0006 & 0.03 & 0.74  \\ 
$1\sigma$ & 0.0013 & 0.00030 & 0.013 & 0.0050 & 0.45 & 0.021 & 3.2 & 0.0026 & 0.048 & 0.56  \\ 
\tableline
\end{tabular}
\vspace{-0.6cm}
\tablenotetext{}{The parameters are: $\Omega_m h^2$ (matter density),
$\Omega_b h^2$ (baryon density), $m_\nu$ (neutrino mass), $Y_{H_e}$ (helium mass fraction),
$z_r$ (reionization redshift), $\theta_s$ (angular sound horizon), $w_x$ (dark energy pressure
to density ratio) and the primordial potential power spectrum is
$k^3P_\Phi(k)=k_f^3P_\Phi(k_f)(k/k_f)^{n_s-1+n_s'\ln{(k/k_f)}}$ with $k_f=0.05{\rm Mpc}^{-1}$.}
\end{center}
\end{table*}

In Figure \ref{fig:planck} we plot the ratio of this systematic offset to the statistical
uncertainty as a function of the maximum multipole considered.
\begin{figure}[!bt]
\vspace{-0.2cm}
\begin{center}
\plotone{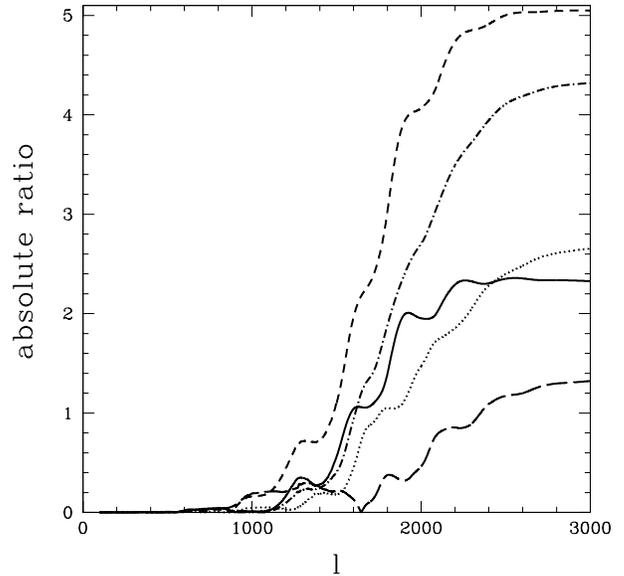}
\vspace{-0.4cm}
\caption{Absolute ratio of bias to 1$\sigma$ for Planck as a function of $\ell_{max}$. Shown are
the parameters that will have a significant bias: $n_s$ (continuous line), $m_\nu$ (dotted),
$Y_{H_e}$ (dashed), $n_s'$ (dot-dashed) and $w_x$ (long-dashed).
\label{fig:planck}} 
\end{center}
\end{figure}
The situation is even worse for a sample variance dominated experiment. 
As we can see in Table~\ref{tab:var}, most of the biases are already important
at $\ell=2000$, thus becoming a problem for parameter estimation.
\begin{table*}[!bt]\small
\begin{center}
\caption{Bias and standard deviation in the parameter estimation for a sample 
variance dominated experiment ($\ell_{\rm max}=2000$)
\label{tab:var}}
\begin{tabular}{l|rrrrrrrrrr}
  & $\Omega_m h^2$ & $\Omega_b h^2$ & $n_s$ & $n_s'$ & $m_\nu$ & $Y_{H_e}$ & $z_r$ &
$\theta_s$ & $\ln{P_\Phi}$ & $w_x$ \\
\hline
bias    &-0.00018 & 0.00030 & -0.030 & -0.0210 & -1.14 & -0.105 & 0.88 & -0.00006 & 0.019 & 0.40\\
$1\sigma$ & 0.00100 & 0.00027 & 0.011 & 0.0049 & 0.43 & 0.018 & 2.87 & 0.0020 & 0.044 & 0.52\\
\hline
\end{tabular}
\end{center}
\end{table*}

The solution is to obtain a model for the patchy power spectrum in terms of a few 
parameters and take them into account when doing a multi-dimensional fit to
the data. Looking at equation (\ref{ppower}) we see that although the amplitude depends
on the reionization model, the shape does not.  Let $C_\ell^1$ be a typical $C_\ell$
from patchy reionization, then any other reionization model can be written as
$C_\ell^f\approx A C_\ell^1$.
There will still be a bias due to the slight differences in shape between the reionization
models (e.g. $C_\ell^R=C_\ell^f-A C_\ell^1$). Table~\ref{tab:sh} shows the ratio,
${\rm bias}/\sigma$, for a cosmic--variance dominated experiment
when considering the amplitude as an extra parameter. We see that the expected bias
is now negligible at the cost of an increase in the statistical uncertainty.

\begin{table*}[!bt]\small
\begin{center}
\caption{Ratio of bias to statistical uncertainty.
\label{tab:sh}}
\begin{tabular}{l|rrrrrrrrrrr}
   & $\Omega_m h^2$ & $\Omega_b h^2$ & $n_s$ & $n_s'$ & $m_\nu$ & $Y_{H_e}$ & $z_r$ &
$\theta_s$ & $\ln{P_\Phi}$ & $w_x$ & A \\
\hline
${bias/\sigma_n}$ &0.027 & 0.027 & -0.060 & -0.083 & -0.030 & -0.077 & -0.008 & -0.031 & -0.008 & 0.007 & 0.011\\
${\sigma_n/\sigma_o}$ & 1.0 & 1.4 & 1.1 & 1.0 & 2.4 & 1.0 & 1.0 & 1.0 & 1.1 & 1.4\\
\hline
\end{tabular}
\vspace{-0.6cm}
\tablenotetext{}{Values are for a sample variance 
dominated experiment ($\ell_{\rm max}=3000$) assuming the patchy signal has a fixed shape. Also shown
is the ratio of the expected $1\sigma$ errors ($\sigma_n$) to the previous values without 
the extra parameter ($\sigma_o$).}
\end{center}
\end{table*}

The dominant source of the increased statistical uncertainty is the remaining uncertainty
in the amplitude of the patchy power spectrum.  This uncertainty can be reduced by
experiments sensitive to $l > 3000$ where the patchy power spectrum dominates,
such as considered in the previous Section.\\ \\

\section{Impact on Lensing Potential Reconstruction}
\label{sec:lens}

To understand the secondary confusion related to patchy reionization in
lensing reconstruction with CMB data, we follow suggested approaches in
the literature \citep{hu02b,cooray03} and modify the noise calculation related
to lensing reconstruction to
include the kSZ contribution. In the case of temperature, the variance of
quadratic statistics, with $\hat{\cmb}^2(\vecl)$ as the filtered
version of the squared temperature (see, for example, \citealt{cooray03}),
 leads to a reconstructed lensing potential power spectrum of
\bea \langle \hat{\cmb}^2(\vecl) \hat{\cmb}^2(\vecl') \rangle = (2\pi)^2
\delta_D(\vecl+\vecl') \left[C_l^{\phi\phi}+N_l^{\phi\phi}\right] \, ,
\eea
where the noise is
\begin{equation}
[N_l^{\phi\phi}]^{-1} =  \int \frac{d^2\vecl_1}{(2\pi)^2} \frac{[\vecl
\cdot \vecl_1 C_{l_1}^P + \vecl \cdot (\vecl-\vecl_1)
C_{|\vecl-\vecl_1|}^P]^2}
{2C_{l_1}^\tot C_{|\vecl-\vecl_1|}^\tot}\, .
\end{equation}
Here, $C_l^P$ is the power spectrum of unlensed primordial temperature
fluctuations, 
while, following equation (\ref{eq:cltot}), $C_l^{\tot}$ accounts for all
contributions to the  temperature power spectrum including noise.
Instead of point sources included in equation (\ref{eq:cltot}), here, we include foreground
contributions related to secondary effects, mainly
thermal (tSZ) and kinetic (kSZ) SZ contributions.  

In Figure~\ref{fig:lens}, we summarize our results where we show the
deflection angle power spectrum, $C_l^{dd} = l(l+1)C_l^{\phi \phi}$
and related noise  contributions. In addition to temperature, 
we have also considered
lensing reconstruction from polarization data and have concentrated on the
quadratic combination related to E and B-mode maps. This combination
provides the best estimator for lensing reconstruction from CMB data if
the signal-to-noise is sufficiently high \citep{hu02b}. 
In calculating these noise curves, we assume temperature
and polarization maps with noise rms of 2 $\mu$K in each
beam--size pixel, and a beam size of 3$'$.
Such sensitivities may be achieved by future ground or space-based
missions.  

The thermal SZ contribution boosts the noise by an order
of magnitude from the case with noise alone. This is, however, not a
significant source of worry since the SZ thermal contribution can
be removed from thermal CMB maps based on the frequency spectrum of the SZ
contribution in multifrequency data. The source of worry will then
be the kSZ signal, including the contribution from patchy reionization.
We show the level of expected noise following our
calculation related to Figure \ref{fig:ksz}. The noise is a factor of a
few
higher than the case of detector noise alone.  

Equation (29) assumes that the patchy power spectrum is perfectly known.
Uncertainties in the amplitude and shape will further increase the noise
in the lensing potential reconstruction, though probably not by much
since measurements at $l \sim 5000$ combined with modeling (such as 
described in our paper) can be used to constrain the amplitude and shape.

While the lensing reconstruction from temperature data is degraded,
this is not the case for polarization. As shown in Figure
\ref{fig:lens}, the noise level including patchy reionization is the
same as when only detector noise is included.  Even for a perfect
polarization map with no detector noise, the reconstruction with
polarization is not significantly affected since the level of
foreground noise is significantly below the cosmic variance limit of
the E-mode power and below the cosmic variance of lensed scalars in
the B-mode power. Thus, polarization observations provide a confusion
free way to extract lensing information and to study primordial
anisotropy contributions.

In addition to foreground noise variance, given by their power spectra,
the foreground secondary effects can also be correlated with lensing
potentials. While this leads to an additional noise contribution, we have
found this  to be significantly smaller than the Gaussian noise shown in
Fig.~\ref{fig:ksz}, since  the quadratic maps  are prefiltered to maximize
lensing reconstruction  such that other higher order correlations  are reduced
\citep{cooray03}. Another secondary source of noise is the
non-Gaussianity of foregrounds themselves. In the case of kSZ related
to patchy reionization, we expect this contribution to be negligible since
at high redshifts clustering is linear and on large scales we will be averaging over the 
contributions from many patches.

\begin{figure}[!bt]
\begin{center}
\includegraphics[width=0.36\textwidth,angle=-90]{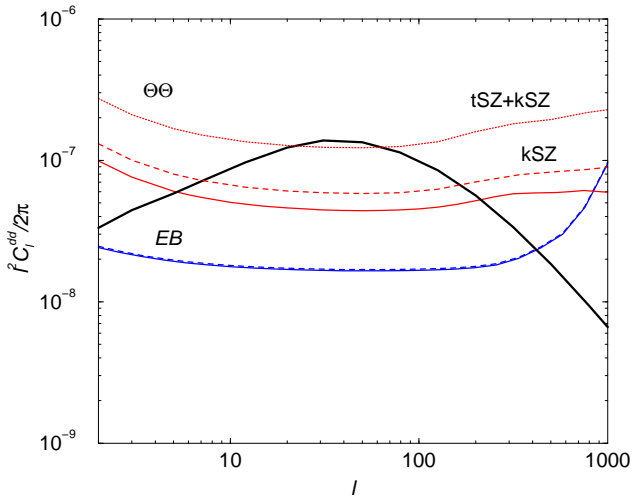}
\caption{The deflection angle power spectrum (black thick solid line)
and the noise variance for its reconstruction with different
assumptions.
In solid red, we show the noise related to reconstruction with temperature
maps while in blue we show the same using polarization data (based on
the quadratic combination of E mode with B mode).
The dotted line shows the reconstruction noise with temperature in the
presence of SZ thermal effect. This secondary contribution can be
removed in multi frequency data using the SZ spectrum. The resulting
lensing reconstruction is then affected by the kSZ signal from
patchy reionization (red dashed line). The blue dashed line shows the same
noise in the presence of kSZ polarization contribution. The latter
is significantly smaller and leads to the same noise when the polarization
from patchy reionization is ignored.
\label{fig:lens}}
\end{center}
\end{figure}

\section{Summary}
\label{sec:concl}

The kSZ power spectrum is important pheno\-me\-no\-lo\-gically in two ways: 1)
as a conta\-minant of the primary/len\-sing CMB power spectrum and 2) as
an interesting probe of the epoch of reionization.  We have shown
that, if neglected, the patchy contribution
can lead to statistically significant biases in the cosmological
parameters estimated from Planck data.  Higher--resolution
observations (such as by ACT, APEX, SPT
\footnote{ACT - Atacama Cosmology Telescope
(http://www.hep.upenn.edu/\~{}angelica/act/act.html),
APEX - Atacama Pathfinder EXperiment (http://bolo.berkeley.edu/apexsz),
SPT - South Pole Telescope (http://astro.uchicago.edu/ spt)}) 
could be even more
strongly affected.  We have proposed a simple solution, which is to
phenomenologically parameterize the patchy power spectrum.  Fortunately
for this purpose, as we have seen analytically, although the amplitude in the relevant l range is
highly uncertain, the shapes are not. For the models we have
considered, extending our cosmological parameter set to include one more
parameter (the amplitude of the patchy power spectrum) eliminates any
significant bias (even for an all-sky cosmic-variance limited
experiment to $\ell=3000$).

In addition to parameter estimation, high--resolution temperature and
polarization anisotropies are affected by gravitational lensing.  CMB
maps can be used to reconstruct the lensing potential, whose auto and
cross power spectra (cross with CMB maps and cosmic shear lensing
potential maps) are sensitive to the dark energy properties and
neutrino masses. We have also investigated the degradation of lensing
potential reconstruction and conclude that lensing studies with
temperature data alone will be affected by fluctuations related to
patchy reionization while polarization data will not be affected.
Thus parameter estimation and lensing--potential reconstruction from
polarization are much less affected by reionization than is the case
for temperature anisotropies.

The kSZ power spectrum can also give us some information about the
reionization process. At $l \ga 3000$ and at frequencies near the null
of the thermal SZ effect, the kSZ reionization power spectrum will be
the dominant source of fluctuation power on the sky, after cleaning
out point sources.  SZ survey instruments such as ACT, SPT, APEX and
CARMA\footnote{Combined Array for Research in Millimeter-wave
Astronomy (http://www.mmarray.org)} with the ability to clean out the
thermal SZ signal based on its spectral dependence will be able
to make highly accurate measurements of the kSZ power spectrum. We have
calculated the expected power spectrum errors for a nominal observation, including
modeling of residual point source contamination. 
If $\tau$ becomes tightly constrained, this type of experiment should allow us to 
determine the mean effective bias and thus provide important information
about the sources of reionization.

In this paper we have only calculated power spectra. We expect that
these two-point functions are the most important statistics, both for
consideration of patchy reionization as a signal and as a contaminant
(see \citealt{castro03} for a discussion of non-Gaussianity from homogeneous
reionization).
At $\ell \la 10^4$, we expect the signal to be approximately Gaussian due
to the large number of patches along each line of sight.  
Non--Gaussian properties may also be of interest and have been studied by
\citet{gnedin02}, though at very small scales ($\ell \sim 10^5$).

\acknowledgments{
We thank Manoj Kaplinghat and Yong-Seon Song
for useful conversations and assistance with
the $C_\ell^P$ derivatives. 
LK and MGS were supported by NASA grant NAG5-11098. AC was supported by
Sherman Fairchild foundation and DOE DE-FG03-92-EX40701.  
CPM is supported by NASA grant NAG5-12173, a Cottrell
Scholars Award from the Research Corporation and an Alfred P. Sloan
fellowship.
}

\bibliography{reion}

\end{document}